%% file: acl_latex.tex
\documentclass[11pt]{article}

\usepackage[preprint]{acl}

\usepackage{times}
\usepackage{latexsym}

\usepackage[T1]{fontenc}

\usepackage[utf8]{inputenc}

\usepackage{microtype}

\usepackage{inconsolata}

\usepackage{graphicx}
\usepackage[most]{tcolorbox}
\usepackage[capitalize]{cleveref}
\usepackage{enumitem}
\usepackage{xcolor}
\usepackage{graphicx}
\usepackage{multirow}
\usepackage{tabularx}
\usepackage{makecell}
\usepackage[table]{xcolor}
\usepackage{booktabs}
\usepackage{arydshln}

\newcommand{\tablestyle}[2]{\setlength{\tabcolsep}{#1}\renewcommand{\arraystretch}{#2}\centering\footnotesize}
\newlength\savewidth\newcommand\shline{\noalign{\global\savewidth\arrayrulewidth
		\global\arrayrulewidth .8pt}\hline\noalign{\global\arrayrulewidth\savewidth}}		
\newcommand\scline[1]{\noalign{\global\savewidth\arrayrulewidth
		\global\arrayrulewidth .8pt}\cline{#1}\noalign{\global\arrayrulewidth\savewidth}}

\definecolor{lightgray}{gray}{0.95}
\definecolor{LIGHT_BLUE}{HTML}{e5f4ff}

\tcbset{
  aibox/.style={
    top=10pt,
    colback=white,
    colframe=black,
    colbacktitle=black,
    enhanced,
    center,
    attach boxed title to top left={yshift=-0.1in,xshift=0.15in},
    boxed title style={boxrule=0pt,colframe=white,},
  }
}
\newtcolorbox{AIbox}[2][]{aibox, title=#2,#1}

\usepackage{amssymb}

\title{TagLLM: A Fine-Grained Tag Generation Approach for Note Recommendation}

\author{
 \textbf{Zhijian Chen\textsuperscript{1}},
 \textbf{Likai Wang\textsuperscript{2,}\thanks{Corresponding author.}},
 \textbf{Lei Chen\textsuperscript{2}},
 \textbf{Yaguang Dou\textsuperscript{2}},
\\
 \textbf{Jialiang Shi\textsuperscript{2}},
 \textbf{Tian Qi\textsuperscript{2}},
 \textbf{Dongdong Hao\textsuperscript{2}},
 \textbf{Mengying Lu\textsuperscript{3}},
 \textbf{Cheng Ye\textsuperscript{2}},
 \textbf{Chao Wei\textsuperscript{2}}
\\
 \textsuperscript{1}Department of Computer Science and Technology, Tongji University, Shanghai, China
\\
 \textsuperscript{2}Shanghai Dewu Information Group Co. Ltd., Shanghai, China
\\
 \textsuperscript{3}Tsinghua Shenzhen International Graduate School,  Tsinghua University, Beijing, China
\\
 \texttt{2533981@tongji.edu.cn}, \texttt{wanglikai@dewu.com}
}

\begin{document}
\maketitle

\input{sec/0_abstract}

\input{sec/1_intro}

\input{sec/2_related_work}

\input{sec/3_method}

\input{sec/4_experiment}

\input{sec/5_conclusion}

\input{sec/6_limit}

\bibliography{custom}

\input{sec/appendix}

\end{document}

%% file: sec/0_abstract.tex
\begin{abstract}

Large Language Models (LLMs) have shown promising potential in E-commerce community recommendation. While LLMs and Multimodal LLMs (MLLMs) are widely used to encode notes into implicit embeddings, leveraging their generative capabilities to represent notes with interpretable tags remains unexplored. In the field of tag generation, traditional close-ended methods heavily rely on the design of tag pools, while existing open-ended methods applied directly to note recommendations face two limitations: (1) MLLMs lack guidance during generation, resulting in redundant tags that fail to capture user interests; (2) The generated tags are often coarse and lack fine-grained representation of notes, interfering with downstream recommendations. To address these limitations, we propose TagLLM, a fine-grained tag generation method for note recommendation. TagLLM captures user interests across note categories through a User Interest Handbook and constructs fine-grained tag data using multimodal CoT Extraction. A Tag Knowledge Distillation method is developed to equip small models with competitive generation capabilities, enhancing inference efficiency. In online A/B test, TagLLM increases average view duration per user by 0.31\%, average interactions per user by 0.96\%, and page view click-through rate in cold-start scenario by 32.37\%, demonstrating its effectiveness.

\end{abstract}

%% file: sec/1_intro.tex
\begin{figure}[!t]
\centering
\includegraphics[width=1\linewidth]{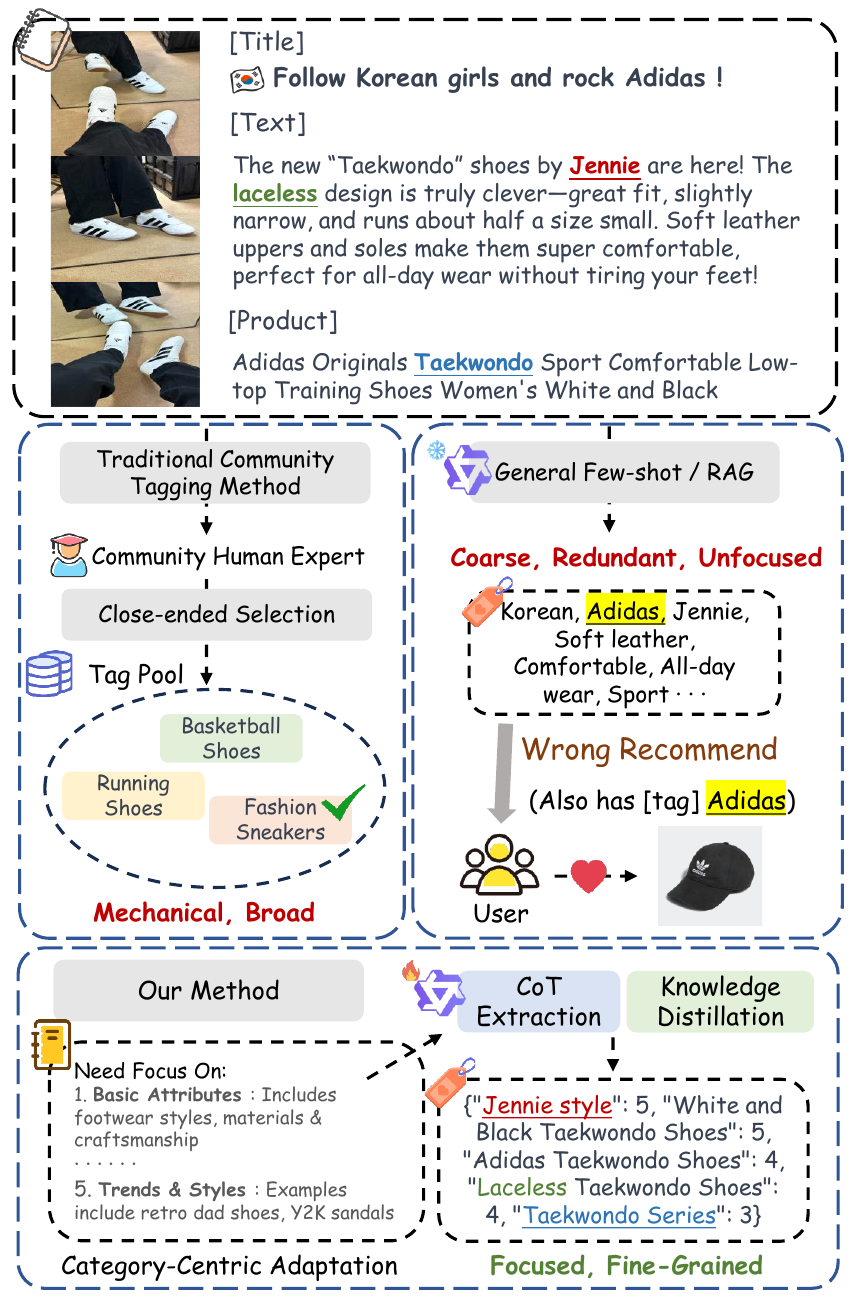}
\vspace{-5mm}
\caption{\textbf{Our Motivation for TagLLM.} Traditional tag generation methods are close-ended, making them difficult to scale and heavily reliant on the tag pool design. Recent LLM-based methods lack guidance, resulting in coarse and unfocused tags. Therefore, we propose TagLLM, a fine-grained generation method that is better suited for note recommendation. }
\label{fig:motivation}
\end{figure}

\section{Introduction}
\label{sec:intro}

With the rapid growth of social media, community culture is increasingly prevalent in E-commerce. Within communities, users share their purchase experiences and recommend products through notes, accelerating product information dissemination and enhancing their sense of belonging to the platform. Community recommendations aim to uncover potential interests of the users, suggesting suitable notes to improve their platform experience. Effective community recommendations can boost the willingness of users to engage with the platform, making it a key focus of research in the E-commerce field.

In community recommendation, Large Language Models (LLMs) for Recommendation have shown their promising potential. Currently, the mainstream approach leverages LLMs or Multimodal LLMs (MLLMs) as encoders to compress notes into embeddings \cite{zhang2024notellm, zhang2025notellm2} for downstream recommendation tasks. While effective and gradually adopted in the industry, this paradigm relies on self-supervised latent representations, which inherently lacks transparency and suffers from unstable supervisory signals. In contrast, using MLLMs to generate explicit tags for notes and applying them remains largely unexplored. In E-commerce communities, tags are key elements that represent notes, summarizing and condensing the core information of notes into a few brief words. Compared to embeddings, tags offer greater control and interpretability when applied to recommendation tasks.

In the field of tag generation, traditional approaches are generally closed-ended. Early methods \cite{efron2010HFB, bougouin2013topicrank} relied on community experts to manually select tags or employed collaborative filtering to assign tags based on co-occurrence probabilities \cite{zangerle2011-tags}. Deep learning-based approaches \cite{zhang2017coattn, kou2018hashtag, zeng2018topic,  hachaj2020vdnn-arm} typically formulate tag generation as a classification task, utilizing CNNs \cite{gong2016attn-cnn} or LSTMs \cite{li2016tab-lstm} to extract features and predict tags from a fixed candidate set. These close-end methods are rigidly constrained by predefined tag pools, making them difficult to scale and unable to capture the differences between diverse notes.



Recently, the powerful generative capabilities of MLLMs have provided a foundation for open-ended tag generation methods. Leveraging MLLMs' prior knowledge and understanding of note content, the content and quantity of tags are no longer fixed, offering greater diversity. Previous studies often rely on MLLM Few-shot Learning \cite{tan2024llm-mhr} or RAG-based \cite{fan2024right} inference methods for tag generation. While these approaches effectively capture comprehensive information from notes, applying them directly to note recommendation has two limitations: (1) MLLMs lack guidance during tag generation, leading to redundant tags that lack focus and fail to reflect dimensions of interest to community users; (2) The generated tags are often coarse and lack fine-grained representation of the notes, which interferes with downstream recommendation tasks. As shown in \cref{fig:motivation}, after coarse-grained generation, both shoes and baseball caps are tagged with "Adidas", failing to truly reflect user interests.

To address these limitations, we introduce \textbf{TagLLM}, a fine-grained tag generation method for note recommendation. TagLLM begins by categorizing user interests through the creation of a User Interest Handbook. Focusing on key dimensions from the Handbook, we design a multimodal CoT Extraction pipeline to construct tag data. CoT Extraction merges low-information tags and ranks them by importance, ensuring the generation of focused and fine-grained tags. To enhance efficiency, a Tag Knowledge Distillation method is developed to enable small-parameter models to achieve competitive tag generation performance through a two-stage training strategy. Finally, the generated tags are transformed into specific features and applied in online A/B test.

In summary, our contributions are three-fold:

\begin{enumerate}[label={\bf {{$\bullet$}}},,leftmargin=*,topsep=0.5ex,itemsep=-0.5ex,partopsep=0.75ex,parsep=0.75ex,partopsep=0pt,wide,labelindent=0pt]
    \item To the best of our knowledge, we propose TagLLM, the first fine-grained tag generation method for note recommendation. TagLLM summarizes user-focused interests based on note categories and creates a Handbook to guide tag generation. Focusing on key dimensions from the Handbook, a CoT Extraction pipeline is developed to construct fine-grained tag data.
    \item We design a Tag Knowledge Distillation method, featuring a two-stage training process and an MLLM-as-a-judge evaluation strategy. This approach enables smaller models to achieve competitive fine-grained tag generation capabilities.
    \item We apply the fine-grained tags as features for downstream recommendation tasks. In the online A/B test, the average view duration per user increases by 0.31\%, the average interactions per user rises by 0.96\%, and page view click-through rate in cold-start scenario grows by 32.37\%.
\end{enumerate}



%% file: sec/2_related_work.tex
\section{Related Work}
\label{sec:related_work}

\begin{figure*}[!t]
\centering
\includegraphics[width=1\linewidth]{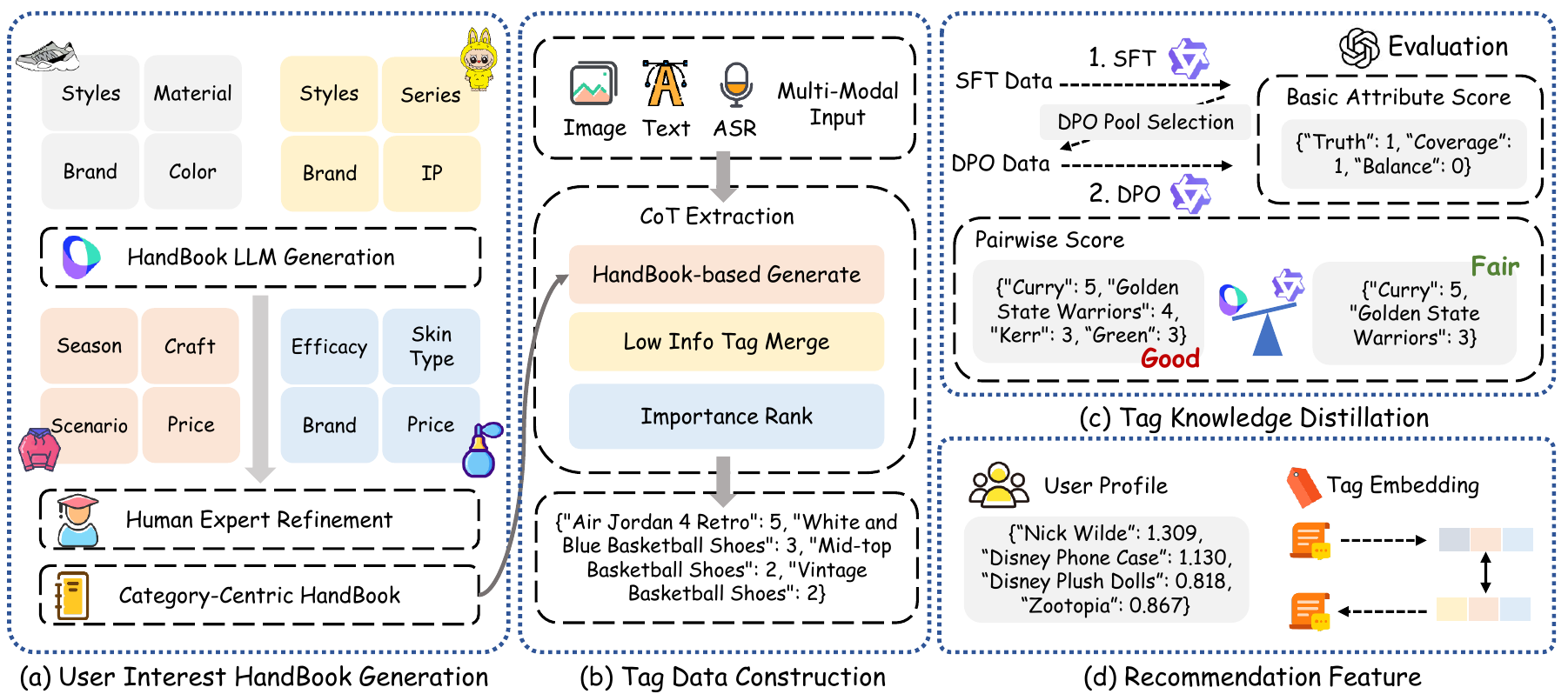}
\vspace{-3mm}
\caption{\textbf{The Overall Pipeline of TagLLM.} First, a User Interest Handbook is generated based on different note categories to reflect user interest. Based on the Handbook, we design a CoT Extraction method to construct tag training data. Next, a Tag Knowledge Distillation method is employed to distill the tag generation capabilities into smaller models. Finally, the tags generated by small models are converted into specific features for recommendations.}
\label{fig:overall_pipeline}
\end{figure*}

\subsection{LLMs for Recommendation}

Current research integrates LLMs into recommendation tasks through three main approaches \cite{zhao2024recommender,lin2025can}. The first approach leverages LLM world knowledge for data enhancement \cite{liu2023first,mysore2023large, zhai2023KP4SR, xi2024towards,wei2024llmrec,dang2025mllmrec, luo2026llama4rec}, struggling with generation quality and train-test alignment. The second approach performs direct recommendations through specific prompts \cite{geng2022p5, liu2023chatgptgoodrecommender, huang2025recommenderaiagent} or fine-tuning \cite{bao2023tallrec, wang2024recmind, yu2025thinkrec, zhang2025recommendation}. While methods like MLLM-MSR \cite{ye2025MLLM-MSR} improve interpretability, context length limits complicate their industrial deployment. The third method uses LLMs as encoders to generate item embeddings \cite{zhang2024notellm,zhong2025speeder, zhang2026persrec}. Although models like NoteLLM-2 \cite{zhang2025notellm2} improve multimodal representation, they inherently lack transparency due to latent embeddings. In contrast, TagLLM proposes a novel tag-feature recommendation method to achieve more controllable and refined recommendations.

\begin{figure*}[!t]
\centering
\includegraphics[width=1\linewidth]{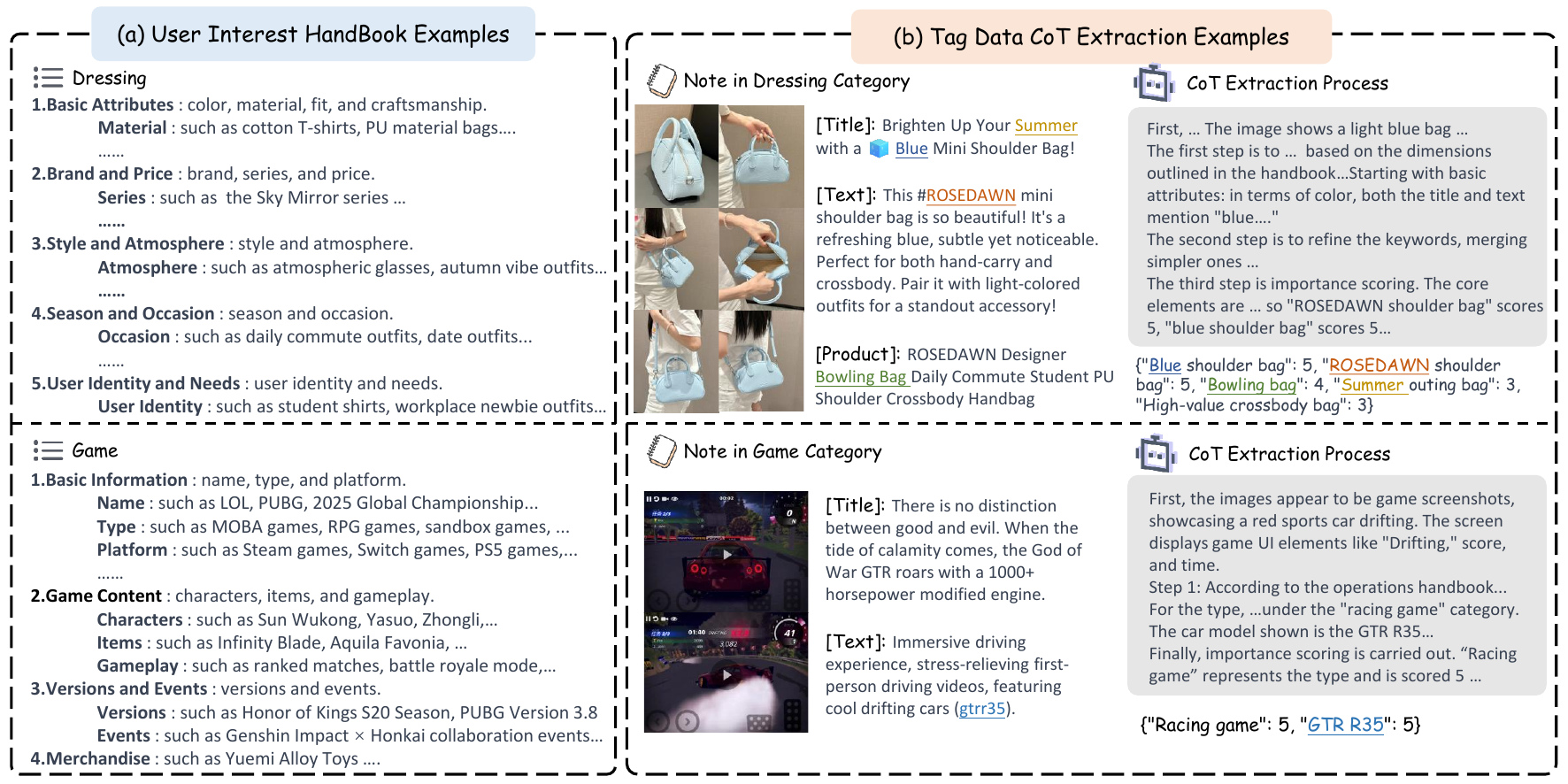}
\vspace{-3mm}
\caption{\textbf{Representative Examples of User Interest Handbook (left) and Tag Data Construction (right).} Through LLM expansion and expert refinement, the guidance in the handbook is fine-grained, covering basic category attributes and user interests. With the three-step CoT Extraction process, LLMs are able to generate high-quality tags based on the guidance.}
\label{fig:represent_example}
\end{figure*}

\subsection{Tag Generation}

In social media, tag generation is a crucial method and can be categorized into three main approaches. The first is the extractive method \cite{efron2010HFB,zangerle2011-tags,zhang2016keyphrase, zhang2018encoding}, which primarily extracts key phrases from text or retrieves tags by identifying similar notes. The second is the classification method \cite{gong2016attn-cnn, li2016tab-lstm, zhang2017coattn, zeng2018topic, hachaj2020vdnn-arm}, which extracts note features using neural networks and feeds them into downstream classification. However, as close-ended methods, both extractive and classification approaches are unable to dynamically represent note features and depend heavily on the quality of the tag pool. The third is the generative method \cite{wang2019topicaware, wang2019microblog, fan2024right, tan2024llm-mhr, xu2025buzz}, which utilizes the powerful generative capabilities and world knowledge of LLMs to generate tags in open-ended domains. However, previous work has focused solely on the representation of notes, leading to redundant and unfocused tags that fail to reflect user interests or meet the needs of community recommendations. TagLLM aims to address these shortcomings and has demonstrated effectiveness in industrial recommendations.


%% file: sec/3_method.tex
\section{Method}
\label{sec:method}

~\cref{fig:overall_pipeline} illustrates the overall pipeline of TagLLM, which consists of four components: User Interest Handbook Generation, Tag Data Construction, Tag Knowledge Distillation, and Recommendation Feature Conversion. In this section, we will introduce each component in detail.

\subsection{User Interest HandBook Generation}

To ensure focused tag generation, we first need to define the user interest dimensions. In the community of E-commerce platforms, notes are typically organized into basic categories (e.g., Dressing, Gaming, Shoes). Given the broad scope of content, it is challenging to apply uniform interest dimensions across all categories. Therefore, we define them on a category-by-category basis, using a two-step approach: LLM-based expansion followed by expert refinement. First, we use the following prompt template to expand potential user interest points in each category:

\begin{tcolorbox}[colback=white, colbacktitle=gray, title=Prompt1: HandBook Generation]
Assume that you are an expert in community operations. Currently, the notes in the community are divided into \emph{<category\_num>} categories. For the notes in the \emph{<category\_name>}, what dimensions of words can serve as content keywords?
\end{tcolorbox}

Next, experts refine and consolidate these points based on the platform's goals to form the final User Interest Handbook. ~\cref{fig:represent_example} presents the handbooks for two categories, Dressing and Gaming, which comprehensively cover the basic attributes of the category content and user interests.

\subsection{Tag Data Construction}

Based on the handbooks for different categories, we employ the CoT Extraction method to explore the multimodal information within notes, which serves as training data for distilling small models.

\noindent\textbf{Multi-Modal Input.} To handle the multimodal nature of notes, we process three key modalities. For visual content, we extract images from the notes and sample video frames at a fixed rate. A maximum number of images is set to ensure stable context length. For text content, we aggregate the title, body, and any associated product descriptions. For audio content, such as voiceovers in videos, we employ SenseVoice \cite{an2024SenseVoice} with LLM-based post-correction to transcribe speech into text.

\noindent\textbf{CoT Extraction.} The CoT (Chain-of-Thought) prompt helps LLMs think step by step, improving task performance while enhancing interpretability. The fine-grained tag generation in CoT Extraction involves three main steps: Handbook-Based Generate, Low-Info Tag Merge, and Importance Rank.

First, we input the handbook along with formatting requirements into the LLM, enabling it to generate user-interest tags based on dimensions from the handbook. The prompt template is as follows:

\begin{tcolorbox}[colback=white, colbacktitle=gray, title=Prompt2: HandBook-based Generate]
\textbf{[Step1]} Please understand the core content of the note, and refer to the key dimensions in the handbook to generate tags.

\dotfill

\textbf{Handbook:} \emph{<Handbook\_Content>} \\
\textbf{Formatting:} \emph{<Style\_Standard>}  
\end{tcolorbox}

Next, we refine the initially generated tags by merging low-information tags using the following prompt template:

\begin{tcolorbox}[colback=white, colbacktitle=gray, title=Prompt3: Low Info Merge]
\textbf{[Step2]} Refine the tags by merging low-information tags and deleting tags that are irrelevant or trivial. If there are inconsistencies between the images and text, discard the tags extracted from the images.   
\end{tcolorbox}

Finally, we rank the refined tags based on importance to make them more focused and reflective of the core content of the note. The prompt template is as follows:

\begin{tcolorbox}[colback=white, colbacktitle=gray, title=Prompt4: Importance Rank]
\textbf{[Step3]} Rank the tags by importance. For each tag, assign an importance score (1-5) based on the core content of the note. A score of 5 indicates the highest importance, and 1 indicates the lowest. 
\end{tcolorbox}

~\cref{fig:represent_example} illustrates examples of Tag Data Construction and the CoT Extraction process. These tags are fine-grained, focused, and effectively reflect the core content of the notes and the users' interests.

\subsection{Tag Knowledge Distillation}

To reduce resource consumption and improve online inference efficiency, we propose the Tag Knowledge Distillation method to distill smaller models, consisting of two training stages and an MLLM-as-a-judge Evaluation Strategy.

\noindent\textbf{Evaluation Strategy.} Since tag generation is an open-ended task, we use the MLLM-as-a-judge approach to evaluate the performance of model distillation. Following WildBench \cite{lin2024wildbench} and Creation-MMBench \cite{fang2025creation-mmbench}, our evaluation consists of two forms: Basic Attribute Scoring and Pairwise Comparison. 

In Basic Attribute Scoring, the judging model evaluates the model across three aspects: Truth, Coverage, and Importance, defined as follows:
\begin{enumerate}[label={\bf {{$\bullet$}}},,leftmargin=*,topsep=0.5ex,itemsep=-0.5ex,partopsep=0.75ex,parsep=0.75ex,partopsep=0pt,wide,labelindent=0pt]
    \item Truth: Assess whether the tags are accurate and realistic, without hallucinations or fabrications.
    \item Coverage: Check whether the tags align with the dimensions defined in the handbook.
    \item Importance: Determine whether the tags reflect the core content of the notes.
\end{enumerate}

Each dimension is scored with 0 or 1. The \textbf{Basic Attribute Score} is the sum of these scores.

In Pairwise Comparison, we generate reference answers for the Eval Set notes using doubao-1.6-thinking. The output of the evaluated model is denoted as A, while the reference answer is B. The judging model compares the two and selects from the set \{A\verb|>>|B, A\verb|>|B, A=B, A\verb|<|B, A\verb|<<|B\}. To facilitate further computation, numerical values are assigned to the results:  \{A\verb|>>|B = +2, A\verb|>|B = +1, A=B = 0, A\verb|<|B = -1, A\verb|<<|B = -2\}. The \textbf{Pairwise Score} is calculated as the average of all scores. To mitigate positional bias in the MLLM-as-a-judge approach, we conduct a Dual Evaluation, swapping the positions of the responses. A detailed evaluation prompt is provided in ~\cref{sec:appendix_d}.

\input{tables/dataset}



\noindent\textbf{Supervised Fine-Tuning (SFT).} In the first stage, we utilize the SFT Set to fine-tune the base model via the standard next-token prediction objective. This process aligns the model's output distribution with the ground-truth labels, establishing a fundamental capability for tag generation and yielding the SFT-trained model.

\noindent\textbf{Direct Preference Optimization (DPO).} In the second stage, we further refine the SFT-trained model using a separate set of notes, referred to as the DPO Pool. To construct preference pairs, we first perform inference on this pool using the SFT-trained model. Subsequently, we conduct the same Pairwise Comparison process as described in the Evaluation Strategy to assess the generated tags against the reference answers. Instances where the generated tags are judged inferior to the references are selected for DPO training, with the reference answer serving as the positive sample and the generated tag as the negative sample.

To verify the effectiveness of each stage, both the intermediate SFT model and the final DPO-refined model are evaluated on the Eval Set.

\input{tables/main_table}

\subsection{Recommendation Feature}

The fine-grained tags of notes are organized into specific features for recommendation, comprising two types: User Profile and Tag Embedding.

\noindent\textbf{User Profile.} For a user $u$, his historical interaction sequence with notes is denoted as $N=\{n_1, n_2, \ldots, n_k\}$. For a note $n_i$, it has a set of fine-grained tags $T=\{t_{i1}, t_{i2}, \ldots, t_{ij}\}$. Based on user $u$'s interactions with note $n_i$, we score the tag set $T$, for instance, by adding $+x\%$ for a click, $+y\%$ for a like, and $+z\%$ for a favorite. This collection of scored tags is then used to construct the user profile, capturing his historical preferences.

\noindent\textbf{Tag Embedding.} For a note $n_i$ with a set of fine-grained tags $T=\{t_{i1}, t_{i2}, \ldots, t_{ij}\}$, we concatenate all tags and use a pretrained encoder to map the tag sequence into a specific latent space as the feature of the note.







%% file: tables/dataset.tex
\begin{table}[t]
\centering
\tablestyle{4pt}{1.2}
\resizebox{\linewidth}{!}{%

\begin{tabular}{c|c|c|c|c|c}
\shline
{\bf DataSets} & {\makecell{\bf Total \\ \bf Num}} & 
{\makecell{\bf Category \\ \bf Num}} & {\makecell{\bf Avg \\ \bf Image \\ \bf Num}}
& {\makecell{\bf Avg \\ \bf Text \\ \bf Length}} & {\makecell{\bf Avg \\ \bf Asr \\ \bf Length}}
\\ \shline

SFT Set & 119,318  & 18 & 4.67 & 92.97 & 144.27 \\
DPO Pool & 19,976 & 18 & 4.73 & 95.43 & 166.31\\
Eval Set & 19,984 & 18 & 4.79 & 103.84  & 173.27\\
\shline

\end{tabular}
}
\caption{\textbf{Statistics of the Experimental Datasets.} }
\vspace{-3mm}
\label{tab: dataset}
\end{table}

%% file: tables/main_table.tex
\begin{table*}[]
\centering
\small 
\resizebox{\textwidth}{!}{
\tablestyle{6pt}{1.2}
\begin{tabular}{l|cccc|cccccc}
\shline
\multirow{2}{*}{\bf Model} &  \multicolumn{4}{c|}{\bf Basic Attribute Score $\uparrow$} &  \multicolumn{6}{c}{\bf Pairwise Score $\uparrow$} \\  \cline{2-11} 
 &  \textbf{Total} & Truth.  & Cover. & Import. & \textbf{Total} & Shoes & Dress & Toys & Beauty & Digit \\ \shline

\rowcolor{lightgray}
\multicolumn{11}{c}{\emph{Base MLLMs}}  \\ \shline

InternVL3-2B & 1.935 & 0.651 & 0.531 & 0.752 & -1.380 & -1.795 & -1.747 & -1.481 & -1.622 & -1.397 \\
InternVL3.5-2B & 1.982 & 0.698 & 0.625 & 0.660 & -1.310 & -1.736 & -1.607 & -1.463 & -1.565 & -1.337 \\
Qwen3-VL-2B-Instruct & 1.412 & 0.510 & 0.396 & 0.501 & -1.657 & -1.862 & -1.952 & -1.715 & -1.837 & -1.712 \\
Qwen3-VL-4B-Instruct & 2.432 & 0.809 & 0.781 & 0.842 & -0.637 & -0.607 & -0.791 & -0.780 & -0.715 & -0.378 \\
\shline

\rowcolor{lightgray}
\multicolumn{11}{c}{\emph{SFT Only}}  \\ \shline

InternVL3-2B & 2.596 & 0.929 & 0.925 & 0.742 & -0.292 & -0.266 & -0.239 & -0.289 & -0.375 & -0.283\\
InternVL3.5-2B & 2.596 & 0.944 & 0.932 & 0.720 & -0.271 & -0.246 & -0.193 & -0.243 & -0.278 & -0.196\\
InternVL3.5-4B & 2.635 & 0.958 & 0.944 & 0.733 & -0.181 & -0.135 & -0.151 & -0.143 & -0.176 & -0.186\\
Qwen3-VL-2B-Instruct & 2.635 & 0.957 & 0.945 & 0.734 & -0.216 & -0.183 & -0.157 & -0.188 & -0.216 & -0.218 \\
Qwen3-VL-4B-Instruct & 2.657 & 0.973 & 0.963 & 0.721 & -0.100 & -0.064 & -0.054 & -0.042 & -0.081 & -0.061 \\
\shline

\rowcolor{lightgray}
\multicolumn{11}{c}{\emph{Tag Knowledge Distillation}}  \\ \shline

InternVL3-2B & 2.654 & 0.945 & 0.958 & 0.750 & -0.341 & -0.012 & -0.302 & -0.370 & -0.420 & -0.376\\
InternVL3.5-2B & 2.636 & 0.902 & 0.938 & 0.796 & -0.156 & -0.083 & -0.164 & -0.219 & -0.239 & -0.150\\
InternVL3.5-4B & 2.669 & 0.944 & 0.954 & 0.771 & -0.057 & 0.150 & -0.055 & -0.092 & -0.112 & -0.020\\
Qwen3-VL-2B-Instruct & 2.650 & 0.941 & 0.950 & 0.758 & -0.126 & -0.131 & -0.101 & -0.102 & -0.163 & -0.137\\
\rowcolor{LIGHT_BLUE}
\textbf{Qwen3-VL-4B-Instruct} & \textbf{2.678} & 0.962 & 0.959 & 0.757 & \textbf{-0.029} & -0.018 & -0.033 & -0.019 & -0.047 & -0.008\\
\shline

\end{tabular}
}
\caption{\textbf{Distillation Performance of MLLMs on Dataset.} Truth., Cover. and Import. stand for the three dimensions in Basic Attribute Scoring: Truth, Coverage and Importance. Shoes, Dress, Toys, Beauty and Digit are the main note categories on the E-commerce platform.}
\label{tab:main-results}
\end{table*}

%% file: sec/4_experiment.tex
\section{Experiment}
\label{sec:experiment}

\subsection{Dataset and Experiment Setting}

\noindent\textbf{Dataset.} We conduct distillation experiments on the note dataset of Dewu, a large-scale E-commerce platform. The SFT Set, DPO Pool and Eval Set all contain the same 18 categories (see ~\cref{sec:appendix_a}). To prevent data leakage, we ensure that data across the three sets is non-overlapping. ~\cref{tab: dataset} provides detailed statistics of the dataset.

\noindent\textbf{Setting.} We select five small models from the InternVL3 \cite{zhu2025internvl3}, InternVL3.5 \cite{wang2025internvl3_5}, and Qwen3-VL-Instruct \cite{bai2025qwen3vltechnicalreport} for distillation experiments. For SFT stage, the learning rate and epochs are set to 1e-4 and 3. For DPO stage, they are set to 5e-6 and 1. Doubao-1.6-thinking is used as the judge model for evaluation. To ensure stability, the temperature for all models is set to 0 during inference.

\subsection{Distillation Performance}

~\cref{tab:main-results} summarizes the distillation performance of various MLLMs on the note dataset, with the following key findings:

\noindent\textbf{Base MLLMs.} Base MLLMs struggle to perform well in fine-grained tag generation tasks, showing poor results on both Basic Attribute Score and Pairwise Score. Likely due to their smaller parameter size, 2B models have difficulties following the guidance in the Handbook and understanding the semantics in notes, resulting in poor accuracy and coverage, as well as a significant gap in pairwise comparisons. In contrast, Qwen3-VL-4B demonstrates strong foundational capabilities, effectively following instructions and completing the generation tasks.

\noindent \textbf{Two-Stage Training.} The distillation results validate the effectiveness of the two-stage training process. After the first SFT stage, all models are able to broadly understand the task requirements and complete the tasks satisfactorily, though some gaps remain in comparison with the reference answers. After fine-tuning in the second stage, the models gain a better understanding of the key aspects of note content. Most models show further improvement in Basic Attribute Score and Pairwise Score , with Qwen3-VL-4B-Instruct achieving the best performance across both metrics, demonstrating near-parity with larger parameter models in generation capabilities.

\noindent \textbf{Category-Level Results.} Across the platform’s main note categories, model performance in the Beauty category is generally lower. This could be attributed to the wide scope of the Beauty, which includes cosmetics, skincare products, and makeup tutorials, making it harder for models to capture consistent patterns. In contrast, the Dressing category, characterized by its strong directionality and relatively fixed structure, enables models to quickly understand and achieve better performance.

\subsection{Ablation Study}
\begin{figure}[t]
\centering
\includegraphics[width=1\linewidth]{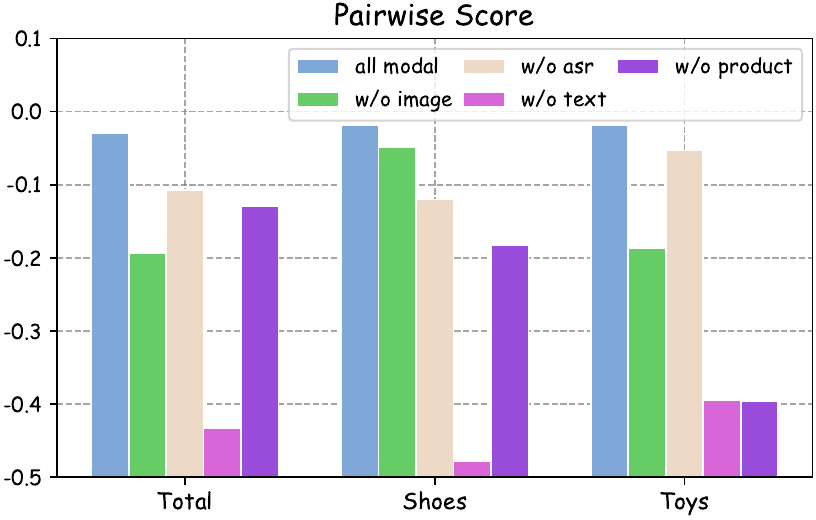}
\vspace{-5mm}
\caption{\textbf{Ablation Study of Input Modalities.} The evaluated model is the distilled Qwen3-VL-4B-Instruct.}
\label{fig:modal_ablation}
\end{figure}

\noindent \textbf{Input Modalities.} ~\cref{fig:modal_ablation} shows the Pairwise Score of the distilled Qwen3-VL-4B-Instruct after removing different input modalities. The body text is the most critical modality in notes. Removing the text results in a significant drop in overall model performance. Visual information also has a notable impact on model effectiveness, although its absence is less pronounced in the Shoes category. This may be due to the higher quality of notes in this category, where rich text and product descriptions compensate for the lack of visual input. In contrast, removing product descriptions in the Toys category leads to a sharp decline in performance, likely because the category is highly product-specific and closely tied to detailed product information.

\subsection{Evaluation Strategy Selection}

The goal of MLLM-as-a-judge is always to achieve higher alignment with human preferences. Therefore, we randomly sample a subset of 100 notes and select three models (InternVL3-2B, InternVL3.5-2B, Qwen3-VL-4B) for tag generation, creating a validation set of 300 questions. We then recruit human experts and select four advanced MLLMs (doubao-1.6-thinking, GPT-4.1, GPT-4o, GPT-5) to judge this set, using MAE and Consistency as metrics to reflect the alignment degree. As shown in ~\cref{tab: model-human-alignment}, doubao-1.6-thinking achieves the best performance in judging consistency for both Basic Attribute Score and Pairwise Score , demonstrating the highest alignment with human preferences. Therefore, we choose it as the judge model for distillation experiments. Specific details are provided in ~\cref{sec:appendix_d}.

\input{tables/human_judge}

\subsection{Failure Case Study}

\begin{figure}[t]
\centering
\includegraphics[width=1\linewidth]{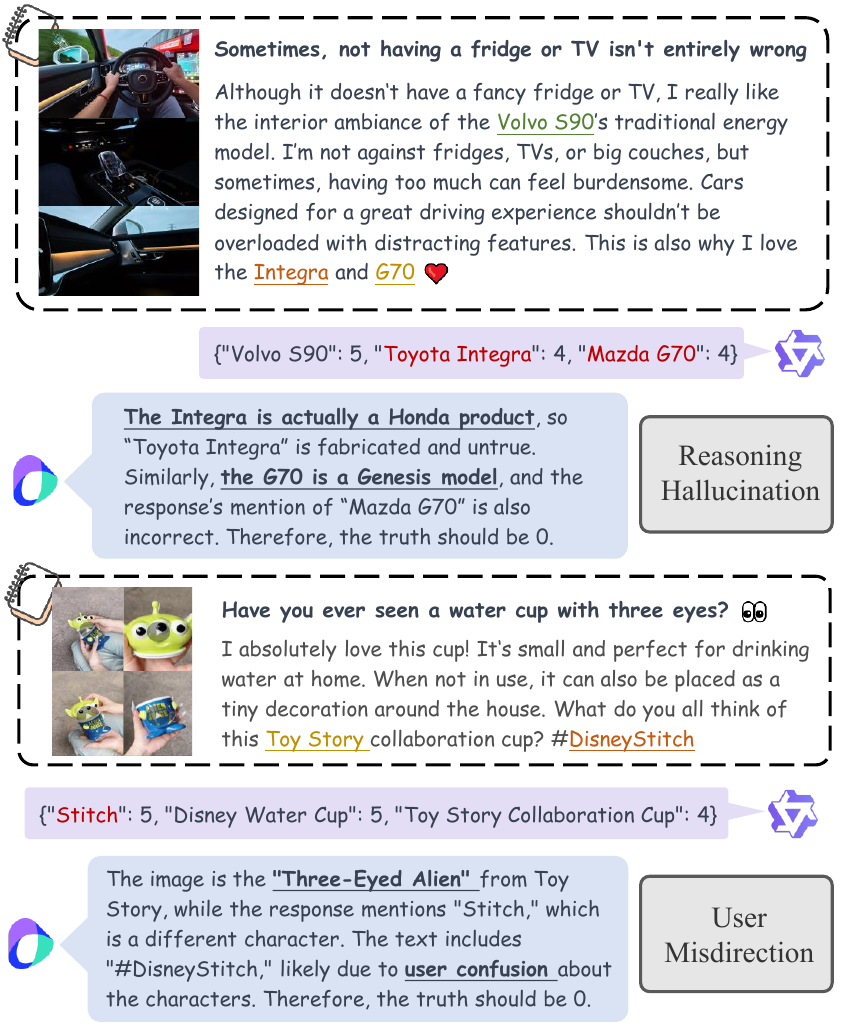}
\vspace{-5mm}
\caption{\textbf{Failure Case Study of TagLLM.} Here are two common types: Reasoning Hallucination and User Misdirection.}
\label{fig:failure_case}
\end{figure}

Although TagLLM is capable of generating high-quality fine-grained tags in most cases, there are certain failure corner cases. ~\cref{fig:failure_case} illustrates two common failure types. The first type is Reasoning Hallucination, which occurs when key information in the notes is incomplete and requires inference to fill in the gaps. The second type is User Misdirection, where user mistakes lead to inconsistencies in the content, misleading the recognition. These types reveal that in certain cases, distilled models lack external knowledge and fail to truly understand the reasoning within the notes. To address these challenges, future efforts could focus on integrating knowledge retrieval tools and applying full-thought process distillation.

\input{tables/A_B_test}

\begin{figure}[t]
\centering
\includegraphics[width=1\linewidth]{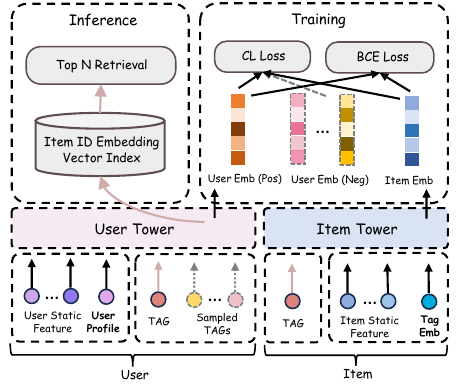}
\vspace{-5mm}
\caption{\textbf{The architecture of the Online Serving Model}, including the training and inference stages.}
\label{fig:online_rec_model}
\end{figure}

\input{tables/cold_start}

\subsection{Online A/B Test}
\noindent \textbf{Online Serving Model.} ~\cref{fig:online_rec_model} illustrates our downstream recommendation architecture. At the input layer, we integrate static user and item features with the recommendation features generated by TagLLM. These inputs are then encoded into embeddings through the User and Item Towers. During training, we jointly optimize these representations using Binary Cross-Entropy (BCE) and Contrastive Learning (CL) losses, capturing both interaction and tag information. For inference, the trained User Tower generates embeddings for each user to perform Top-$K$ retrieval within the Item ID Index. More Details are provided in ~\cref{sec:appendix_e}.

\noindent \textbf{Overall Improvement.} ~\cref{tab:A_B_test} shows the overall performance of TagLLM in the online A/B test. We can observe that fine-grained tags effectively model user-item interactions, leading to significant increases of 0.31\% in AVDU and 0.96\% in AIU.

\noindent \textbf{Cold-start Improvement.}  To evaluate TagLLM in the cold-start scenario, we monitor newly published notes within 24 hours. ~\cref{tab:cold_start} presents cold-start performance across different channels. The results demonstrate that fine-grained tags effectively compensate for the sparse interactions of cold-start notes. Compared to fcandiw, an existing tag-based inverted index method, TagLLM achieves substantial improvements of 32.37\% in PVCTR and 46.91\% in PVIR, validating its effectiveness for cold-start recommendations.

%% file: tables/human_judge.tex
\begin{table}[t]
\centering
\tablestyle{4pt}{1.1}
\resizebox{\linewidth}{!}{%

\begin{tabular}{c|c|cc|cc}
\shline
\multirow{2}{*}{\bf Judger} & \multirow{2}{*}{\bf MLLM} & \multicolumn{2}{c|}{\bf BAS} & \multicolumn{2}{c}{\bf PS} \\ \scline{3-6}
& & \textbf{MAE$\downarrow$} & \textbf{Cons.$\uparrow$} & \textbf{MAE$\downarrow$} & \textbf{Cons.$\uparrow$} \\ \shline

\multirow{4}{*}{\makecell{\textbf{doubao-1.6} \\ \textbf{-thinking}}} & Intern3 & 0.15  & 100.0 & 0.33  & 94.95  \\
& Intern3.5 & 0.14  & 100.0  & 0.36  & 97.00  \\
& Qwen3 & 0.15 & 100.0 & 0.34 & 96.00 \\ 
\rowcolor{lightgray}
\cellcolor{white} & Total & \textbf{0.15} & \textbf{100.0} & \textbf{0.34} & \textbf{95.99} \\ \shline

\multirow{4}{*}{GPT-4.1} & Intern3 & 0.22  & 97.92 & 0.52  & 85.57  \\
& Intern3.5 & 0.16  & 100.0  & 0.43  & 94.85  \\
& Qwen3 & 0.15 & 100.0 & 0.47 & 90.62 \\ 
\rowcolor{lightgray}
\cellcolor{white} & Total & 0.17 & 99.30 & 0.48 & 90.34 \\ \shline

\multirow{4}{*}{GPT-4o} & Intern3 & 0.30  & 92.78 & 0.67  & 82.76  \\
& Intern3.5 & 0.27  & 95.88  & 0.60  & 87.10  \\
& Qwen3 & 0.24 & 95.88 & 0.50 & 87.72 \\ 
\rowcolor{lightgray}
\cellcolor{white} & Total & 0.27 & 94.85 & 0.59 & 85.88 \\ \shline

\multirow{4}{*}{GPT-5} & Intern3 & 0.39  & 93.81 & 0.61  & 92.78  \\
& Intern3.5 & 0.46  & 90.72  & 0.59  & 94.85  \\
& Qwen3 & 0.39 & 92.78 & 0.48 & 92.78 \\ 
\rowcolor{lightgray}
\cellcolor{white} & Total & 0.42 & 92.44 & 0.56 & 93.47 \\ \shline

\end{tabular}
}
\caption{\textbf{The Alignment Between Different Judge Models and Human Preference.} BAS and PS respectively denote Basic Attribute Score and Pairwise Score.}
\label{tab: model-human-alignment}
\end{table}

%% file: tables/A_B_test.tex
\begin{table}[t]
\centering
\tablestyle{4pt}{1.2}
\resizebox{\linewidth}{!}{%

\begin{tabular}{c|ccccc}
\shline
{\bf Metrics} & {AVDU} & {AIU}
& {GMV} & {UVCTR} & {VCU}
\\ \shline

TagLLM & +0.31\%  & +0.96\% & +0.15\% & +0.06\% & +0.21\% \\
\shline

\end{tabular}
}
\caption{\textbf{Overall improvement in Online A/B Test.} AVDU, AIU, GMV, UVCTR and VCU respectively denote Average View Duration per User, Average Interactions per User, Gross Merchandise Volume, Unique Visitor Click-Through Rate and Valid Clicks per User.}
\label{tab:A_B_test}
\end{table}

%% file: tables/cold_start.tex
\begin{table}[t]
\centering
\tablestyle{4pt}{1.2}
\resizebox{\linewidth}{!}{%




\begin{tabular}{c|cc|c|cc}
\shline
{\textbf{Channel}} & {PVCTR}
& {PVIR} & {\textbf{Channel}} & {PVCTR} & {PVIR}
\\ \shline

TagLLM & +32.37\% & +46.91\% & TagLLM- & +31.66\% & +34.03\% \\
\shline

\end{tabular}
}
\caption{\textbf{Improvement in the cold-start scenario.} The tag-based channel fcandiw serves as the baseline. TagLLM- is a simplified variant of TagLLM. PVCTR, PVIR respectively denote Page View Click-Through Rate and Page View Interaction Rate.}
\label{tab:cold_start}
\end{table}

%% file: sec/5_conclusion.tex
\section{Conclusion}
\label{sec:conclusion}

In this paper, we explore a new tag-centric path for LLMs in recommendation. To address the limitations of previous methods, we propose TagLLM , a fine-grained tag generation method for note recommendation. TagLLM leverages a User Interest Handbook to describe user interests across different note categories and utilizes CoT Extraction to construct tag data step by step. A Tag Knowledge Distillation method is designed to transfer generative capabilities to smaller models, improving inference efficiency. Finally, the features are converted into specific formats for online A/B testing , resulting in increases of 0.31\%, 0.96\%, and 32.37\% in AVDU, AIU, and cold-start PVCTR, respectively.

In the future, we will introduce VideoLLM to further enhance video understanding within notes and introduce tools such as knowledge retrieval to mitigate reasoning hallucinations, enabling more accurate tag generation.



%% file: sec/6_limit.tex
\section*{Limitations}

While TagLLM explores a new path for LLMs recommendation and provides high-quality tags in most cases, we acknowledge the following limitations:

\noindent \textbf{Video Understanding.} Currently, TagLLM extracts video information using fixed-frequency frame sampling. While effective in most simple scenarios, it struggles with processing complex and long videos. In the future, advanced long-video understanding methods will be introduced to enhance its capabilities.

\noindent \textbf{Reasoning Hallucination.} Hallucination remains an inevitable challenge in large model reasoning. At present, the model may experience reasoning hallucinations when attempting to complete incomplete information in notes. Moving forward, we plan to integrate knowledge retrieval tools and error correction mechanisms to address this issue.

\noindent \textbf{Strong Logical Reasoning.} Due to user mistakes, notes may sometimes contain contradictory content. The current model could be misled by these inconsistencies, resulting in suboptimal answers. In the future, RLHF based on thought processes will be considered to strengthen the model’s reasoning capabilities.


%% file: sec/appendix.tex
\appendix

\section{Category Description}
\label{sec:appendix_a}

On the Dewu E-commerce platform, notes can be categorized into 18 classes. The category names and descriptions are as follows:

\begin{enumerate}[label={\bf {{$\bullet$}}},,leftmargin=*,topsep=0.5ex,itemsep=-0.5ex,partopsep=0.75ex,parsep=0.75ex,partopsep=0pt,wide,labelindent=0pt]
    \item \textbf{Shoes:} Showcases various types of shoes, including running shoes, fashion shoes, and basketball shoes.
    \item \textbf{Dressing:} Includes exhibitions of clothing, pants, accessories, and outfit recommendations.
    \item \textbf{Sports:} Covers information about various sports and recommendations for related equipment.
    \item \textbf{Beauty:} Features recommendations for makeup, personal care products, and tutorials on makeup techniques.
    \item \textbf{Digit:} Includes recommendations and reviews for digital devices and equipment.
    \item \textbf{Fitness:} Covers fitness tutorials, equipment, weight-loss meals, and related accessories.
    \item \textbf{Home:} Showcases various home products and tutorials on space planning.
    \item \textbf{Toys:} Includes plush toys and alloy models.
    \item \textbf{Food:} Features cooking tutorials and restaurant exploration vlogs.
    \item \textbf{Lifestyle:} Captures various life records and reflections.
    \item \textbf{Entertainment:} Includes information and activities related to entertainment.
    \item \textbf{Education:} Covers various courses and tutorials on life skills.
    \item \textbf{Travel:} Includes travel guides and experience sharing.
    \item \textbf{Media:} Includes information and reflections on books, films, and music.
    \item \textbf{Dance:} Features various dance records and tutorials.
    \item \textbf{Game:} Includes game guides and reviews of related equipment.
    \item \textbf{Pets:} Features pet-related sharing, knowledge, and product recommendations.
    \item \textbf{Arts:} Covers various artistic works and experience sharing.

\end{enumerate}

\section{User Interest HandBook and Tag Data}
\label{sec:appendix_b}
\subsection{CoT Extraction}
In the main text, we introduced the core parts of the three steps in CoT Extraction. A more detailed and complete CoT Extraction Prompt Template is shown in ~\cref{fig:full_cot_extraction_template}. As illustrated, CoT Extraction constrains tag generation through four aspects of Tag Requirements and provides few-shot examples to help the model better understand the task objectives. After completing the core three steps, a fourth step is conducted to review the execution of the prior steps, ensuring the reliability of the tags.

\subsection{More Handbook and Tag Data Examples}
\begin{figure*}[]
\centering
\includegraphics[width=0.95\linewidth]{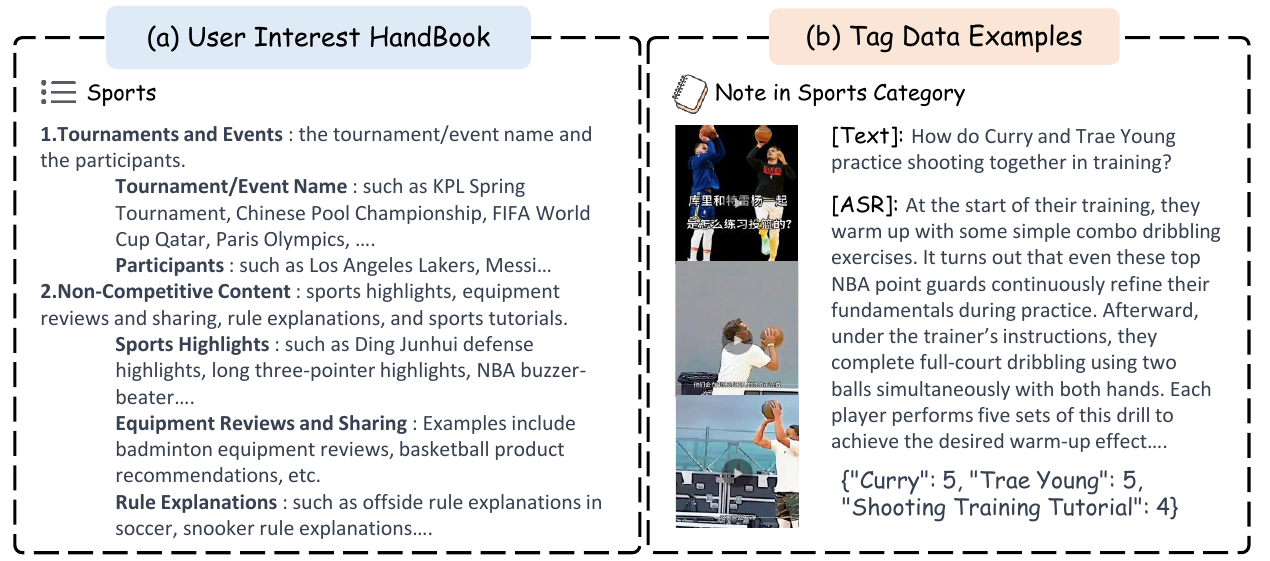}
\vspace{-3mm}
\caption{\textbf{User Interest Handbook and Tag Data Examples of the Sports.}}
\label{fig:appendix_sports}
\end{figure*}

\begin{figure*}[]
\centering
\includegraphics[width=0.95\linewidth]{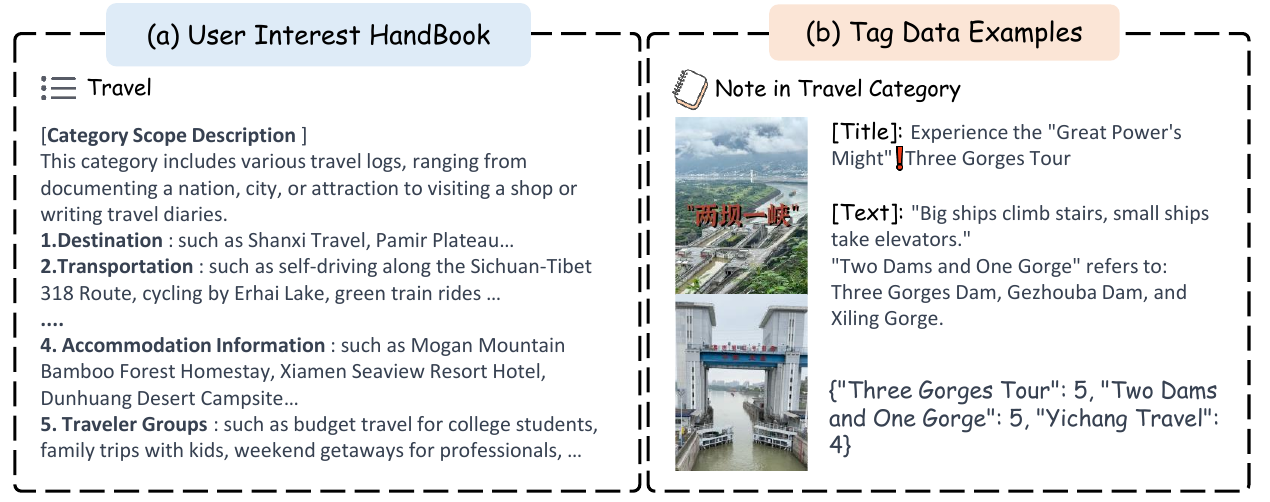}
\vspace{-3mm}
\caption{\textbf{User Interest Handbook and Tag Data Examples of the Travel.}}
\label{fig:appendix_travel}
\end{figure*}

\begin{figure*}[]
\centering
\includegraphics[width=0.95\linewidth]{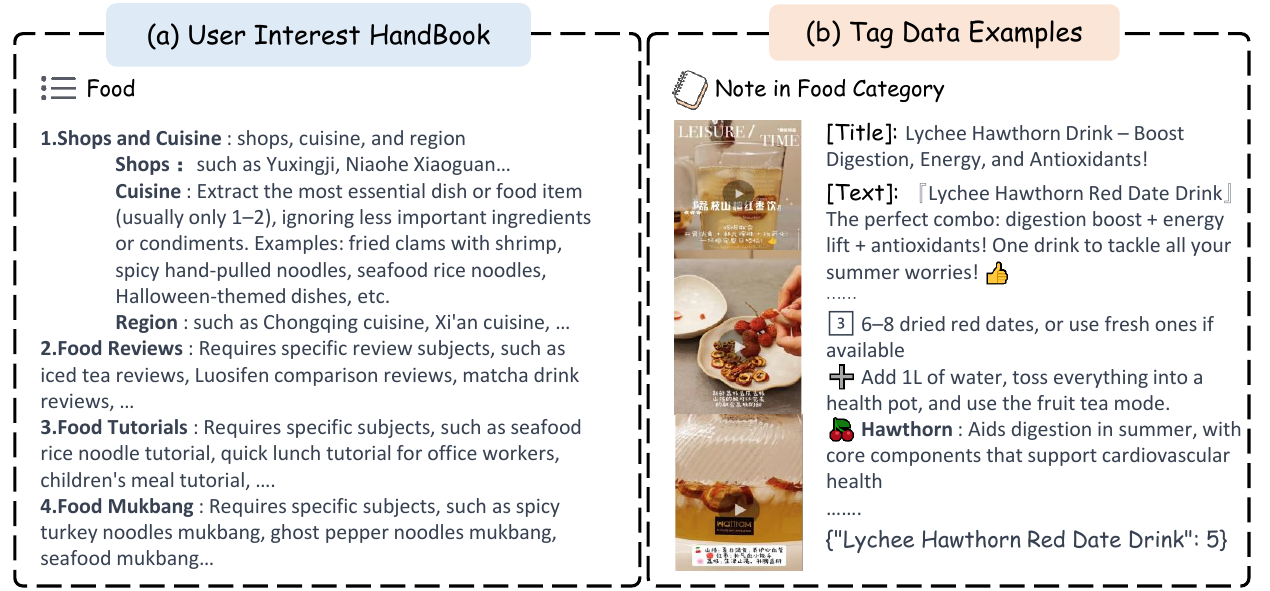}
\vspace{-3mm}
\caption{\textbf{User Interest Handbook and Tag Data Examples of the Food.}}
\label{fig:appendix_food}
\end{figure*}

\begin{figure*}[]
\centering
\includegraphics[width=0.95\linewidth]{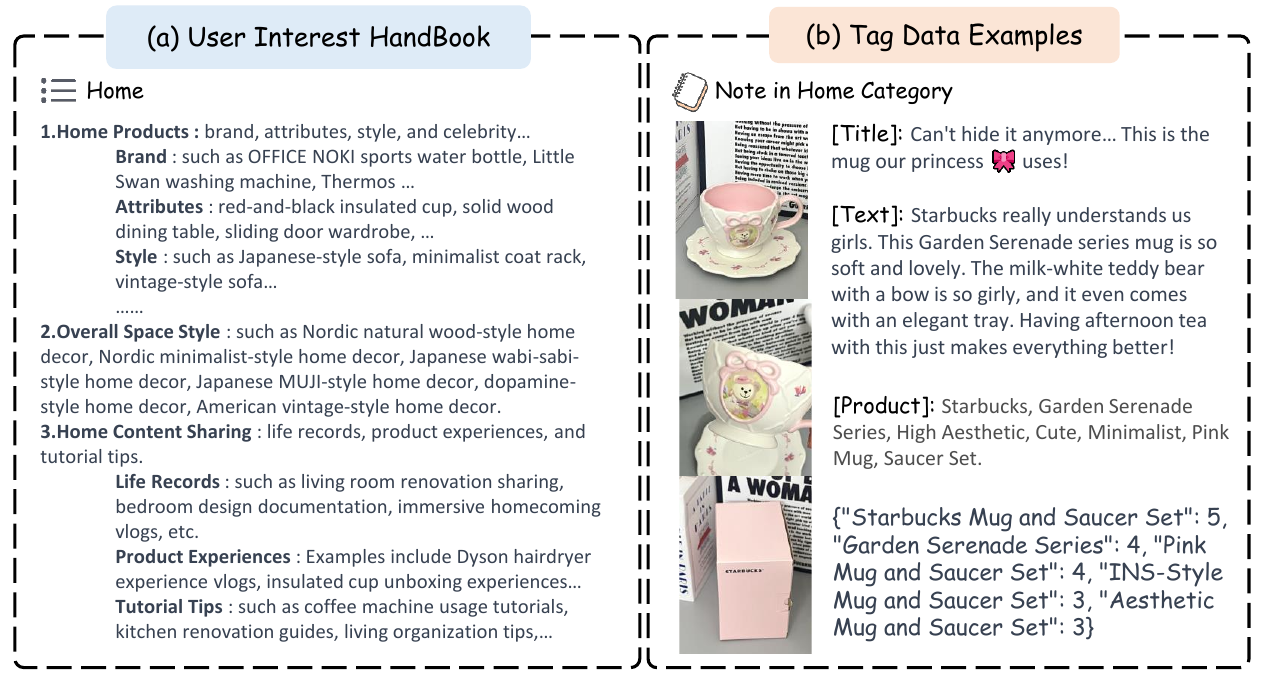}
\vspace{-3mm}
\caption{\textbf{User Interest Handbook and Tag Data Examples of the Home.}}
\label{fig:appendix_home}
\end{figure*}

\begin{figure*}[]
\centering
\includegraphics[width=0.95\linewidth]{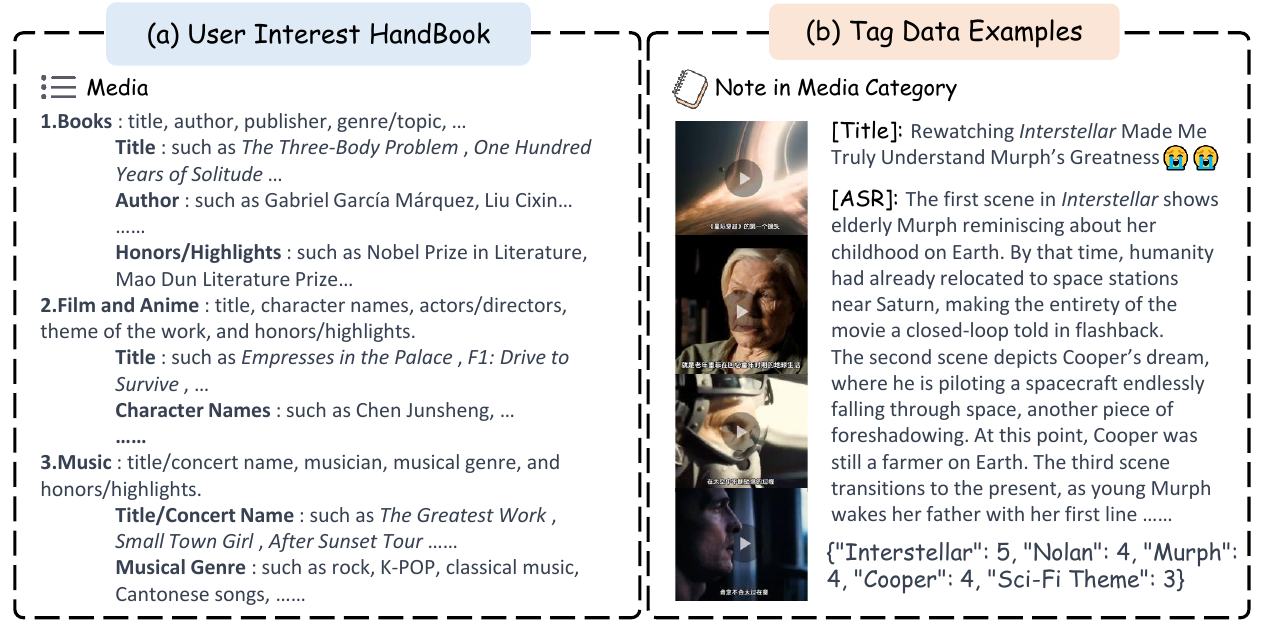}
\vspace{-3mm}
\caption{\textbf{User Interest Handbook and Tag Data Examples of the Media.}}
\label{fig:appendix_media}
\end{figure*}

\begin{figure*}[]
\centering
\includegraphics[width=0.95\linewidth]{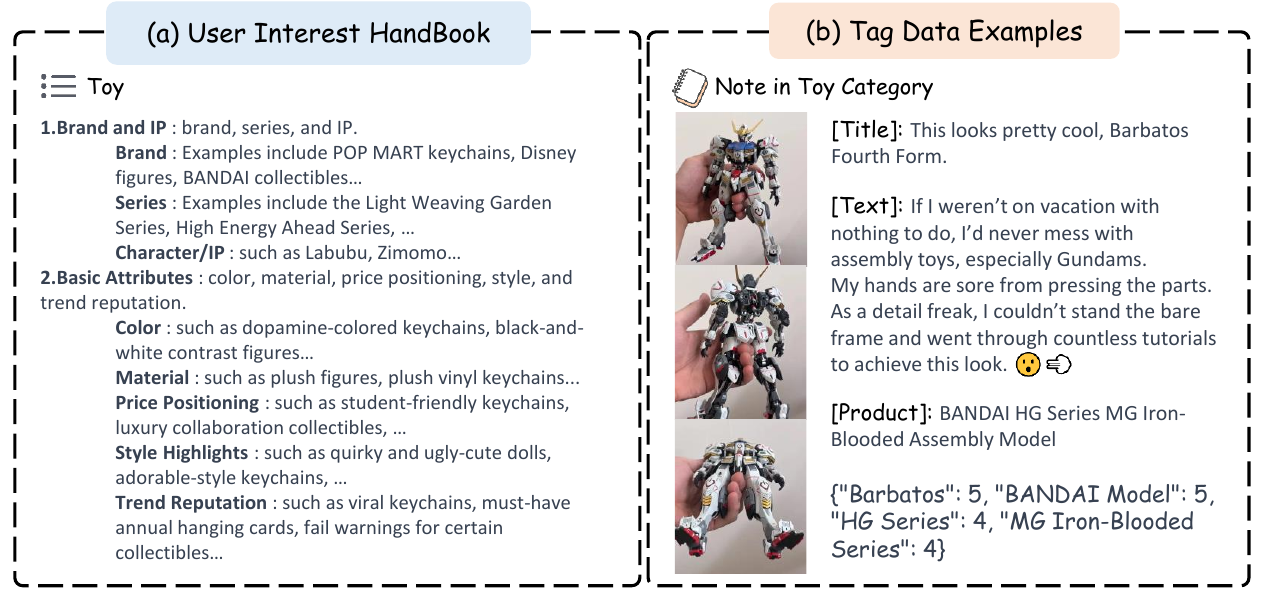}
\vspace{-3mm}
\caption{\textbf{User Interest Handbook and Tag Data Examples of the Toy.}}
\label{fig:appendix_toy}
\end{figure*}

\begin{figure*}[]
\centering
\includegraphics[width=0.95\linewidth]{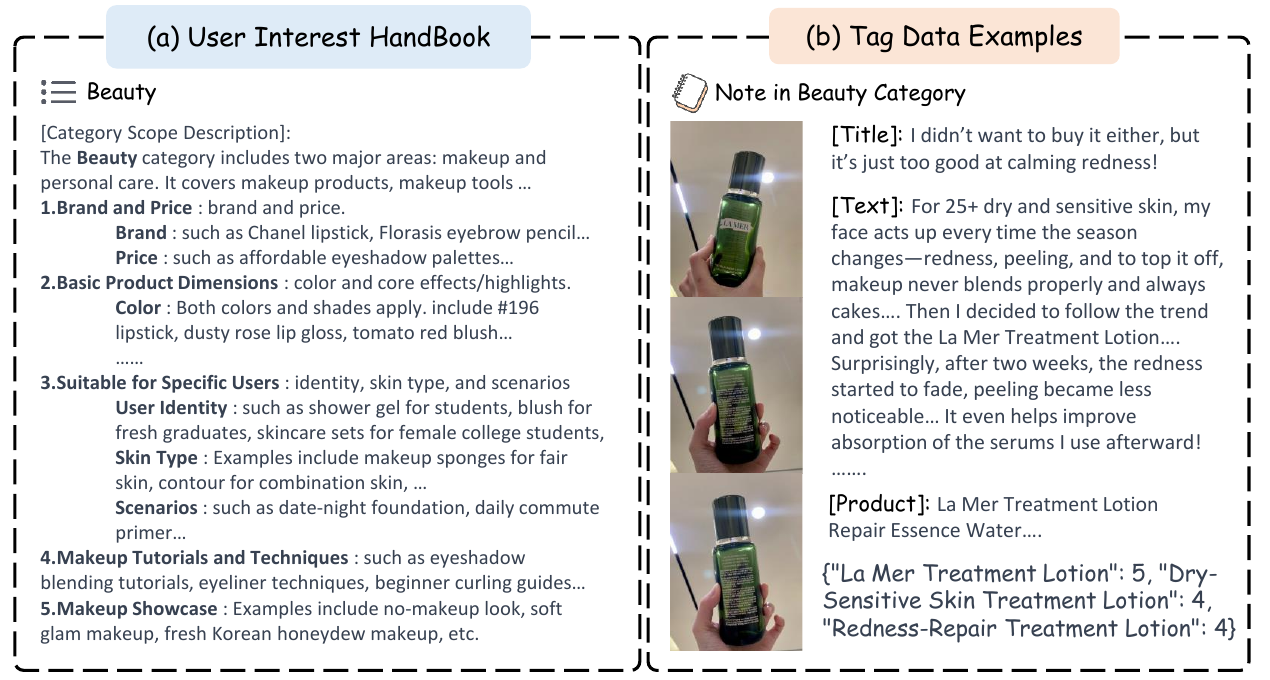}
\vspace{-3mm}
\caption{\textbf{User Interest Handbook and Tag Data Examples of the Beauty.}}
\label{fig:appendix_beauty}
\end{figure*}

\begin{figure*}[]
\centering
\includegraphics[width=0.95\linewidth]{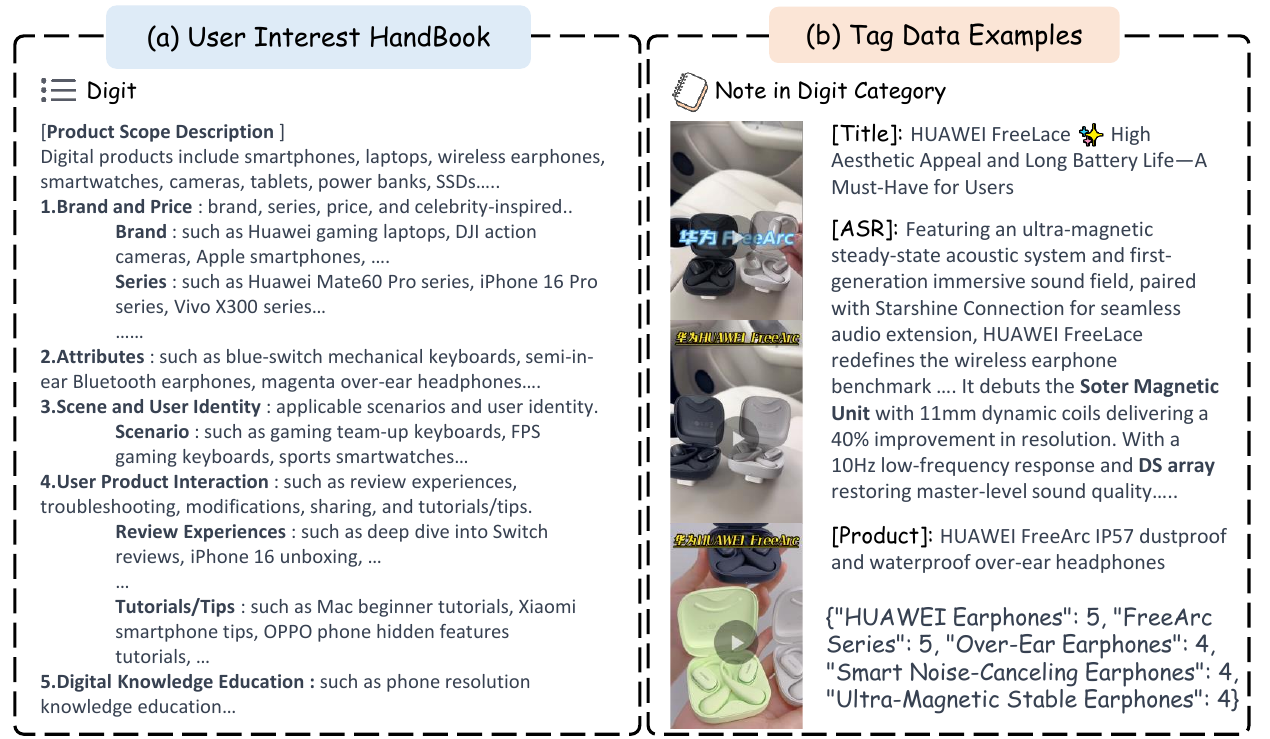}
\vspace{-3mm}
\caption{\textbf{User Interest Handbook and Tag Data Examples of the Digit.}}
\label{fig:appendix_digit}
\end{figure*}

In the main text, we provided Handbook and Tag Data examples of the Dressing and Game categories to demonstrate our overall process. To give readers a more comprehensive understanding of our work, ~\cref{fig:appendix_sports} - ~\cref{fig:appendix_digit} present additional examples of 8 categories.

\section{Tag Knowledge Distillation}
\label{sec:appendix_c}

\input{tables/appendix_pairwise}

~\cref{tab:appendix_pairwise} shows the detailed pairwise judgment results of various MLLMs. It can be observed that Base MLLMs exhibit a significant number of Much Worse cases, leading to poor overall Pairwise Score. After applying Tag Knowledge Distillation, the Much Worse cases are significantly reduced, while the Much Better cases increase to some extent. Among all models, distilled Qwen3-VL-4B-Instruct achieves the best performance in Much Better and Better cases.

\section{Evaluation Strategy and Human Alignment}
\label{sec:appendix_d}

\subsection{Evaluation Prompt Template}

~\cref{fig:evaluation_prompt_bas} and ~\cref{fig:evaluation_prompt_ps} present the evaluation prompt templates we used for Basic Attribute Scoring and Pairwise Comparison. As shown, we incorporated CoT guidance into the evaluation process to ensure the judge model thinks and executes step by step, enhancing the stability of the assessment.

\subsection{Alignment Metrics}

Eq ~\eqref{eq:mae} and ~\eqref{eq:consistency} present the metrics (MAE and Consistency) used to evaluate the degree of alignment. In these equations, $\mathcal{J}$ represents the basic attribute scoring or pairwise comparison results from a specific judging model, while $\mathcal{P}$ denotes the corresponding reference value (average of human ratings).

\begin{equation}
    \textbf{MAE} = \frac{1}{n} \sum_{i=1}^{n} | \mathcal{J}_i - \mathcal{P}_i |
    \label{eq:mae}
\end{equation}

\begin{equation}
    \textbf{Consistency} = \frac{1}{n} \sum_{i=1}^{n} 
    \begin{cases} 
        1, & \text{if } |\mathcal{J}_i - \mathcal{P}_i| \leq 1  \\
        0, & \text{otherwise}
    \end{cases}
    \label{eq:consistency}
\end{equation}

\begin{figure}[t]
\centering
\includegraphics[width=1\linewidth]{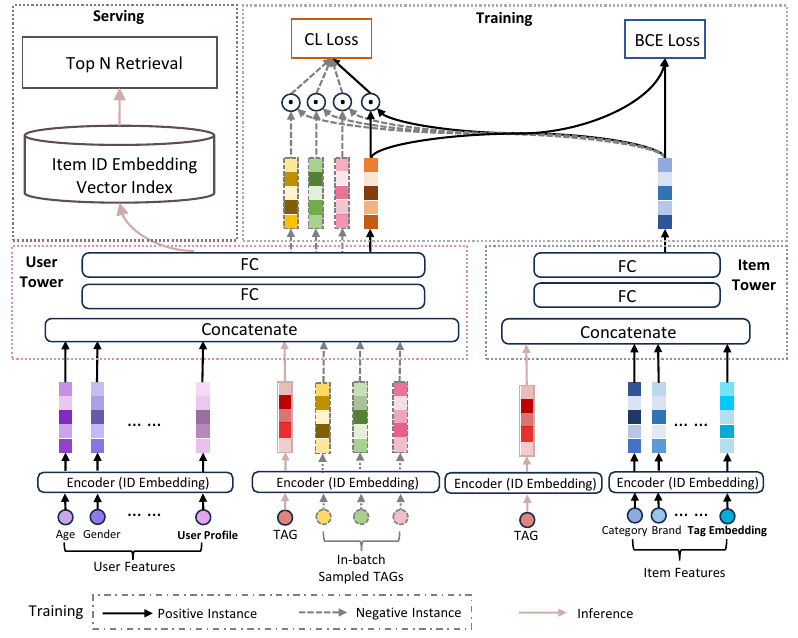}
\vspace{-3mm}
\caption{\textbf{The Details of the Online Serving Model.}}
\label{fig:appendix_online_rec_model}
\end{figure}

\section{Online Serving Model}
\label{sec:appendix_e}

~\cref{fig:appendix_online_rec_model} illustrates the details of the online serving model, which consists of four components: the input layer, the dual-tower architecture, the training stage, and the inference stage.

\subsection{Input Layer}
For the User Tower, we input the user's static features (e.g., age, gender) along with the User Profile generated by TagLLM. These features are mapped to ID embeddings, denoted as $f^u$. We treat the tags from notes clicked by the user as positive samples, and tags randomly sampled within the batch as negative samples. These tags are similarly represented via ID embeddings as $z$.

For the Item Tower, we input the item's static features (e.g., brand, category) and the Tag Embeddings generated by TagLLM. These features are mapped to ID embeddings, denoted as $f^v$. The tags from notes clicked by the user are also represented as $z$ via ID embeddings.

\subsection{User Tower and Item Tower}
In the User Tower, we concatenate the user features $f^u$ and the tag representation $z$, then pass them through two fully connected (FC) layers to obtain:
\begin{equation}
    U(z, f^u) = \text{FC}(\text{FC}(\text{concat}(f^u, z)))
\end{equation}

Similarly, in the Item Tower, we concatenate the item features $f^v$ and the tag representation $z$, then pass them through two fully connected layers to obtain:
\begin{equation}
    V(z, f^v) = \text{FC}(\text{FC}(\text{concat}(f^v, z)))
\end{equation}

\subsection{Training Stage}
During the training stage of the online model, we jointly optimize the user and item representations using Binary Cross-Entropy (BCE) loss and Contrastive Learning (CL) loss.

The BCE loss is used to model the user's click preferences for historical items, formulated as:
\begin{equation}
    \mathcal{L}_{\text{BCE}} = - [y \log(\hat{y}) + (1 - y) \log(1 - \hat{y})]
\end{equation}
where $\hat{y}$ is the predicted click probability for the item.

The CL loss is employed to optimize the joint representation of tags, users, and items. $U(z, f^u)$ and $U(z', f^u)$ denote the user tag representations under positive and negative sample tags respectively, while $V(z, f^v)$ represents the item tag representation under the positive sample tag. The CL loss aims to minimize the representation distance between $U(z, f^u)$ and $V(z, f^v)$, and maximize the distance between $U(z', f^u)$ and $V(z, f^v)$. It is computed as follows:
\begin{equation}
    \mathcal{L}_{\text{CL}} = \frac{e^{\langle U(z, f^u), V(z, f^v) \rangle}}{e^{\langle U(z, f^u), V(z, f^v) \rangle} + \sum_{z' \in \mathbb{Z}} e^{\langle U(z', f^u), V(z, f^v) \rangle}}
\end{equation}

Finally, the jointly optimized loss function is defined as:
\begin{equation}
    \mathcal{L} = \mathcal{L}_{\text{BCE}} + \lambda_1 \mathcal{L}_{\text{CL}} + \lambda_2 \|\Theta\|^2,
\end{equation}
where $\Theta$ is the set of model parameters, and $\lambda_1$ and $\lambda_2$ are hyperparameters.

\subsection{Inference Stage}
In the inference stage, we generate the user representation $\hat{u}$ through the well-trained User Tower, combined with the tags of previously clicked notes. Using $\hat{u}$, we retrieve the top-$K$ candidate items from the item vector index. These items are then merged with the results from other channels for further downstream processing.

\begin{figure*}[p] 
\begin{AIbox}{Full CoT Extraction Prompt Template}
{   
    Assume that you are an expert in operating a short video and graphic content platform, responsible for the \textcolor{brown}{\textless category\textgreater}. You now need to complete the tag generation task for multimedia content in this category. Based on the given content (including images, titles, body text, and product details), summarize and extract key information. \\ \\
    You have a category-specific Handbook recording the dimensions users in this category care about. The Handbook is as follows: \\ \\
    \textbf{HandBook:} \textcolor{brown}{\textless HandBook Content\textgreater}
    \\ \\ 
    You need to work according to the following steps: \\ \\
    \textbf{[Step 1]:} Understand the core content of the multimedia content and generate tags based on the key dimensions outlined in the Handbook. \\
    \textbf{Tag requirements:} \textcolor{brown}{\textless Quantity\textgreater+\textless Part of Speech\textgreater+\textless Format\textgreater+\textless Special\textgreater}
    \\ \\
    \textbf{[Step 2]:} Deeply analyze the content's core ideas. Refine tags generated in Step 1: \\
    $ \bullet $  Merge simple, low-information tags. Delete unrelated, trivial, or colloquial keywords that do not align with the core content. \\
    $ \bullet $  Focus on tag quality over quantity, retaining only keywords most relevant to the content. If the content lacks substance or contains no valid information, the keywords should be left blank. \\
    $ \bullet $  If inconsistencies are found between product content and the main multimedia content, delete keywords extracted from those products. \\
    $ \bullet $  If inconsistencies are found between the image and text information, disregard keywords extracted from the image and prioritize text-based content.
    \\ \\
    \textbf{[Step 3]:} Assign importance scores to the keywords refined in Step 2. \\
    Evaluate each keyword's relevance to the multimedia content and assign an importance score (1-5), where 5 indicates the highest importance and 1 indicates the lowest. Sort keywords and their corresponding importance scores in descending order of importance.
    \\ \\
    \textbf{[Step 4]: }Perform a final review of the refined keywords: \\
    Ensure they meet the above standards and do not include trivial or meaningless tags. Finalize the output. If no tags exist, output an empty string. If tags exist, output them in JSON format. Do not add any additional information beyond the required output.
    \\ \\
    \textbf{Result example:} \textcolor{brown}{\textless Category Few-Shot Example\textgreater}
    \\
    \textbf{Note Content:} \textcolor{brown}{\textless Note\textgreater}
}
\end{AIbox}
\caption{\textbf{The Full Prompt Template of CoT Extraction}}
\label{fig:full_cot_extraction_template}
\end{figure*}

\begin{figure*}[] 
\begin{AIbox}{Evaluation Prompt Template -- Basic Attribute Score}
{   
    Assume you are an expert in operating a short video and graphic content platform, adept at analyzing and evaluating note content. Now, your task is to score an answer for a note tag generation task based on the corresponding scoring criteria. \\

    \textbf{[Step 1]:} Understand the goal and requirements of the note tag generation task (provided in <Note Tag Task Details>). 
    
    \textbf{[Step 2]:} Based on the task’s goal and requirements, review the scoring criteria (provided in <Scoring Criteria>) and the corresponding task response (provided in <Task Response>). If there is no content in <Task Response>, the answer is an empty tag. 
    
    \textbf{[Step 3]:} Evaluate the task response according to the task’s goal and scoring criteria. Represent the score in JSON format, for example: \{'Truth': 1, 'Coverage': 0, 'Importance': 1\}. 
    
    \textbf{[Step 4]:} Verify whether the above three steps have been correctly executed. If accurate, output only the score JSON without any additional information.
    \\
    
    \textbf{Note Tag Task:} \textcolor{brown}{\textless Note Tag Task Details\textgreater}
    \\
    
    \textcolor{blue}{\textless Scoring Criteria\textgreater} \\
    1. \textbf{Truth: } If the tag is empty, assign 1 point. If the tag is non-empty, evaluate whether the tag content aligns with the textual and visual content of the post, ensuring no fabrication or hallucination. Assign 1 point if all tags are accurate; otherwise, 0 points. \\
    2. \textbf{Coverage:} If the tag is empty, evaluate whether the empty tag aligns with the goal and requirements of the task provided in <Note Tag Task Details>. Assign 1 point if correct; otherwise, 0 points. If the tag is non-empty, determine whether all tags fit within the dimensions in the operational handbook outlined in <Note Tag Task Details>. Assign 1 point if they do; otherwise, 0 points. Note that the evaluation of coverage is independent of authenticity. Even if the tag includes fabricated or hallucinatory content, it can still score 1 for coverage if it meets the criteria. \\
    3. \textbf{Importance:} If the tag is empty, assign 0 points. If the tag is non-empty, determine whether the tag content reflects prioritization based on the post content. If the post is content-oriented, the number of tags describing the content should not be fewer than those describing the product; if product-oriented, the number of tags describing the product should not be fewer than those describing the content. Assign 1 point if the condition is met; otherwise, 0 points. \\
    
    \textbf{Tags:} \textcolor{brown}{\textless Task Response \textgreater}
}
\end{AIbox}
\caption{\textbf{The Evaluation Prompt Template of Basic Attribute Scoring}}
\label{fig:evaluation_prompt_bas}
\end{figure*}

\begin{figure*}[] 
\begin{AIbox}{Evaluation Prompt Template -- Pairwise Comparison}
{   
    Assume you are an expert in operating a short video and graphic content platform, proficient in analyzing and evaluating note content. You are tasked with comparing the quality of two responses, A and B, for a note tag generation task and providing the judgment result.\\

    \textbf{[Step 1]:} Understand the goal and requirements of the note tag generation task (provided in <Note Tag Task Details>).
    
    \textbf{[Step 2]:} Based on the task’s goal and requirements, compare the quality of responses A and B (provided in <Task Responses>). \\
    The comparison result must be one of the following five outcomes: \{A\verb|>>|B, A\verb|>|B, A=B, A\verb|<|B, A\verb|<<|B\} (representing A is far better than, better than, roughly equal to, worse than, or far worse than B in quality, respectively). \\
    Note that the result must place A before B, such as "B\verb|>|A" is not acceptable.
    
    \textbf{[Step 3]:} Verify whether the execution of the first two steps is error-free. If correct, output the final result without any additional information.
    \\\\
    \textbf{Note Tag Task:} \textcolor{brown}{\textless Note Tag Task Details\textgreater}
    \\
    \textbf{Criteria:} \textcolor{brown}{\textless Pairwise Criteria\textgreater}
    \\\\
    \textbf{\textless Task Response\textgreater} \\
    \textbf{A:} \textcolor{blue}{\textless Tag\_A\textgreater} \quad\quad \textbf{B:} \textcolor{blue}{\textless Tag\_B\textgreater}
}
\end{AIbox}
\caption{\textbf{The Evaluation Prompt Template of Pairwise Comparison}}
\label{fig:evaluation_prompt_ps}
\end{figure*}

%% file: tables/appendix_pairwise.tex
\begin{table}[]
\centering
\small 
\resizebox{\linewidth}{!}{
\tablestyle{4pt}{1.1}
\begin{tabular}{c|cccccc}
\shline
 \textbf{Model} & \textbf{PS} & \textbf{MB} & \textbf{B} & \textbf{T} & \textbf{W} & \textbf{MW}\\ \shline

\rowcolor{lightgray}
\multicolumn{7}{c}{\emph{Base MLLMs}}  \\ \shline

InternVL3-2B & -1.380 & 634 & 2,346 & 550 & 14,071 & 22,283 \\
InternVL3.5-2B & -1.310 & 686 & 2,340 & 1,957 & 13,852 & 21,067 \\
Qwen3-VL-2B & -1.657 & 305 & 931 & 895 & 7,845 & 29,826 \\
Qwen3-VL-4B & -0.637 & 1,608 & 7,913 & 1,889 & 20,285 & 8,087 \\
\shline

\rowcolor{lightgray}
\multicolumn{7}{c}{\emph{SFT Only}}  \\ \shline

InternVL3-2B & -0.292 & 1,147 & 9,581 & 8,073 & 18,453 & 2,508\\
InternVL3.5-2B & -0.271 & 912 & 10,264 & 8,472 & 17,490 & 2,690\\
InternVL3.5-4B & -0.181 & 1,026 & 11,187 & 9,022 & 16,721 & 1,854\\
Qwen3-VL-2B & -0.216 & 1,076 & 10,705 & 8,799 & 17,047 & 2,199 \\
Qwen3-VL-4B & -0.100 & 990 & 12,294 & 9,877 & 15,137 & 1,558 \\
\shline

\rowcolor{lightgray}
\multicolumn{7}{c}{\emph{Tag Knowledge Distillation}}  \\ \shline

InternVL3-2B & -0.341 & 1,264 & 9,795 & 6,983 & 17,655 & 4,133\\
InternVL3.5-2B & -0.156 & 1,664 & 12,003 & 6,966 & 16,879 & 2,340\\
InternVL3.5-4B & -0.057 & 1,459 & 13,199 & 8,435 & 15,123 & 1,628\\
Qwen3-VL-2B & -0.126 & 1,445 & 11,880 & 8,448 & 16,292 & 1,757\\
Qwen3-VL-4B & -0.029 & 1,490 & 13,073 & 9,125 & 14,857 & 1,167\\
\shline

\end{tabular}
}
\caption{\textbf{The Details of Pairwise Comparision Performance.} PS denotes Pairwise Score. MB, B, T, W, MW stand for Much Better, Better, Tie, Worse, Much Worse.}
\label{tab:appendix_pairwise}
\end{table}